\documentclass[a4paper]{article}

\usepackage{INTERSPEECH2020}
\usepackage{xcolor}
\usepackage{amsmath,graphicx,color,multirow,diagbox,amssymb,threeparttable,mathtools,booktabs,float,siunitx}
\usepackage{cite}

\title{Distortionless Multi-Channel Target Speech Enhancement for Overlapped Speech Recognition}
\name{\em Bo Wu$^{1}$, Meng Yu$^{2}$, Lianwu Chen$^{1}$, Yong Xu$^{2}$, Chao Weng$^{2}$, Dan Su$^{1}$, and Dong Yu$^{2}$}
\address{
  $^1$Tencent AI Lab, Shenzhen, China\\
  $^2$Tencent AI Lab, Bellevue, WA, USA}
\email{\{lambowu, raymondmyu, lianwuchen, lucayongxu, cweng, dansu, dyu\}@tencent.com}

\begin{document}

\maketitle
\begin{abstract}
Speech enhancement techniques based on deep learning have brought significant improvement on speech quality and intelligibility. Nevertheless, a large gain in speech quality measured by objective metrics, such as perceptual evaluation of speech quality (PESQ), does not necessarily lead to improved speech recognition performance due to speech distortion in the enhancement stage. In this paper, a multi-channel dilated convolutional network based frequency domain modeling is presented to enhance target speaker in the far-field, noisy and multi-talker conditions. We study three approaches towards distortionless waveforms for overlapped speech recognition: estimating complex ideal ratio mask with an infinite range, incorporating the fbank loss in a multi-objective learning and finetuning the enhancement model by an acoustic model. Experimental results proved the effectiveness of all three approaches on reducing speech distortions and improving recognition accuracy. Particularly, the jointly tuned enhancement model works very well with other standalone acoustic model on real test data.
 \end{abstract}

\section{Introduction}
In the presence of interfering speakers, the target speech intelligibility is usually degraded in the mixed signal. Such deterioration can severely affect automatic speech recognition (ASR). Although many techniques have been developed for speech enhancement/separation \cite{wang2018supervised, hershey2016deep, isik2016single,luo2018speaker,chen2017deep,yu2017permutation,kolbaek2017multitalker,luo2018tasnet,luo2019conv} and recognition \cite{weng2015deep, barker2015third, wu2017end} under these circumstances, it still remains one of the most challenging problems in ASR. A large gain in speech quality can translate to a negligible improvement in recognition accuracy \cite{bahmaninezhad2019comprehensive,8461639}. The discrepancy in objectives between speech enhancement and recognition results in the performance inconsistency. Compared to the aggressive noise reduction, the subtle speech distortion introduced in the enhancement stage does not affect much on the enhancement loss or evaluation metrics, such as PESQ and signal-to-distortion ratio (SDR). Determined by the training loss and non-linear activations in the neural network, the situation of speech distortion is even worse in low signal-to-noise ratio (SNR) conditions. Such distortion is harmful to ASR. 

To reduce speech distortion in the front-end processing, the study in \cite{gao2016snr} proposes a progressive learning framework by guiding each hidden layer of the deep neural network to learn an intermediate target with gradual signal-to-noise ratio gains explicitly. The work presented in \cite{xu2017multi} imposes additional continuity constraints to alleviate the over-estimate or under-estimate problems in the reconstructed signal. Wang \textit{et al.} overcomes the distortion problem by performing a distortion independent back-end acoustic model \cite{wang2019bridging}. Moreover, jointly modeling the front-end enhancement and back-end acoustic model is another desirable solution for improving recognition accuracy in noisy and multi-talker environments. For example, Chang \textit{et al.} designs a neural sequence-to-sequence architecture for end-to-end multi-channel multi-speaker speech recognition in \cite{chang2019mimo}. Other researchers in \cite{ochiai2017unified,xu2019joint,inproceedings} propose to jointly train a neural beamformer and acoustic model for noise robust ASR. Nevertheless, most of them fail to answer what happens to the enhanced waveforms through the joint training \cite{wu2017end,7403942} or whether the enhancement model fine-tuned by the jointly trained acoustic model still helps recognition on a standalone ASR system \cite{chang2019mimo,narayanan2014improving}.

In this paper, based on our previous work on end-to-end multi-channel convolutional TasNet with short-time Fourier transform (STFT) kernel for target speech enhancement \cite{bahmaninezhad2019comprehensive}, we study and compare three main approaches towards distortionless waveforms for recognition, regarding mask types, target domains and loss functions, respectively.  
The contribution of this paper is three-fold. First, we compare various distortionless approaches in the same setup, which is often separately discussed in different studies. Second, with the joint optimization on magnitude and phase, we find that uncompressed complex ideal ratio mask (cIRM) leads to significant protection on target speech signal \cite{williamson2016complex,williamson2015complex}. Third, thanks to the error back-propagation from the loss of acoustic model \cite{wu2017end,bahdanau2016end}, the speech enhancement model produces distortionless signals, resulting in effective ASR performance in this particular end-to-end joint training setup.  Furthermore, we show that such enhanced signals work well with standalone ASR system on large recorded test sets as well. 

The rest of the paper is organized as follows. In Section \ref{sec:modules}, we recap our direction-aware multi-channel enhancement network. In Section \ref{sec:method}, we present three distortionless methods. We describe our experimental setups and evaluate the effectiveness of the presented approaches in Section \ref{sec:exp}. We conclude this work in Section \ref{con}.

\begin{figure*}[htbp]
        \centering
        \includegraphics[width=\linewidth]{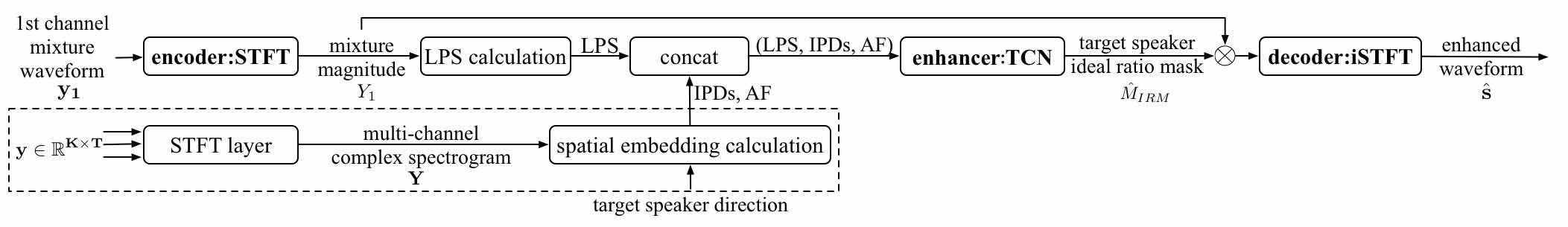}
        \vspace{-0.7cm}
        \caption{Block diagram of the multi-channel target speech enhancement network.}
        \label{fig:tasnet}
\end{figure*}

\section{Multi-Channel Enhancement }
\label{sec:modules}

Figure \ref{fig:tasnet} shows our previous work on direction-aware multi-channel target speech enhancement framework \cite{bahmaninezhad2019comprehensive}, which recovers target speaker's voice from the reverberant, noisy and multi-talker mixed signal. It consists of three major parts: 

\noindent(1) An encoder (a fixed STFT convolution 1-D layer) transforms the input waveform to STFT domain. A reference channel, usually 1st channel waveform $\bf{y}_{1}$ without the loss of generality is transformed to spectral magnitude $Y_{1}$ which is used to compute log-power spectral (LPS) by $\log({Y}_1^2)$. The LPS feature vector is then concatenated with inter-channel phase differences (IPDs) and target speaker-dependent angle feature (AF) \cite{gu2019neural}. IPD feature represents spatial location information \cite{chen2018multi} and is calculated by the phase difference between two channels of complex spectrogram as:
\vspace{-0.25cm}
\begin{equation}
\text{IPD}_{k}(t,f)=\angle{\frac{\mathbf{Y}_{k1}(t,f)}{\mathbf{Y}_{k2}(t,f)}}
\vspace{-0.1cm}
\end{equation}
where $k1$ and $k2$ are two microphones of the $k$-th microphone pair and $K$ is the total number of selected microphone pairs. An angle feature is incorporated as a target speaker bias. This feature was originally introduced in \cite{chen2018multi}, which computes the averaged cosine distance between the target speaker steering vector and IPD on all selected microphone pairs as
\vspace{-0.1cm}
\begin{equation}
\text{AF}(t, f)=\overset{K}{\underset{k=1}{\sum}}
\frac{
\mathbf{e}_{k}(f)
\frac{\mathbf{Y}_{k1}(t,f)}{\mathbf{Y}_{k2}(t,f)}}
{
\left |\mathbf{e}_{k}(f)
\frac{\mathbf{Y}_{k1}(t,f)}{\mathbf{Y}_{k2}(t,f)} \right |
}
\vspace{-0.1cm}
\end{equation}
where $\mathbf{e}_{k}(f)$ is the steering vector coefficient target speaker at frequency \emph{f} with respect to $k$-th microphone pair. As a result, AF indicates if a speaker from a desired direction dominates in each time-frequency bin, which drives the network to extract the target speaker from the mixture.

\noindent(2) An enhancement block estimates the target speaker's ideal ratio mask. A temporal fully-convolutional network (TCN) \cite{luo2019conv} is adopted in the enhancement network which infers the target speaker's ideal ratio mask $\hat{M}_{\text{IRM}}$ activated by ReLu function and $\theta$ is the model parameter:
\vspace{-0.1cm}
\begin{equation}
g(\text{LPS,  IPDs, AF}; \theta)=\hat{M}_{\text{IRM}}
\vspace{-0.1cm}
\end{equation}

\noindent(3) A decoder (a fixed iSTFT convolution 1-D layer) reconstructs waveform. A single-channel enhanced waveform $\hat{\bf{s}}$ is reconstructed from the multiplication between mixture magnitude $Y_{1}$ and target speaker mask $\hat{M}_{\text{IRM}}$ as:
\vspace{-0.1cm}
 \begin{equation}
\hat{\mathbf{s}}=\text{iSTFT}(\hat{M}_{\text{IRM}}\otimes Y_{1}, \varphi)
\vspace{-0.1cm}
\end{equation}
where $\otimes$ is the element-wise product of two operands and $\varphi$ represents the first-channel mixture speech phase.

The scale-invariant signal-to-distortion (SI-SNR) is used as the objective function to optimize the enhancement network which is defined as:
\vspace{-0.1cm}
\begin{equation}
    \text{SI-SNR}:=10\log_{10}\frac
{\left\|\mathbf{s}_{\sf target}\right\|_{2}^{2}}
{\left\|\mathbf{e}_{\sf noise}\right\|_{2}^{2}}
\vspace{-0.1cm}
\end{equation}
where $\mathbf{s}_{\sf target}=\left(\left<\hat{\mathbf{s}}, \mathbf{s}\right>\mathbf{s}\right)/\left\|\mathbf{s}\right\|_{2}^{2}$, $\mathbf{e}_{\sf noise}=\hat{\mathbf{s}}-\mathbf{s}_{\sf target}$, and $\hat{\mathbf{s}}$ and $\mathbf{s}$ are the estimated and reverberant target speech waveforms, respectively. The zero-mean normalization is applied to $\hat{\mathbf{s}}$ and $\mathbf{s}$ for scale invariance. We refer the readers to \cite{bahmaninezhad2019comprehensive} for more details about the implementation of the multi-channel target speech enhancement model.

\section{Distortionless Methods}\label{sec:method}
\begin{figure}[!h]
        \centering
        \includegraphics[width=\linewidth]{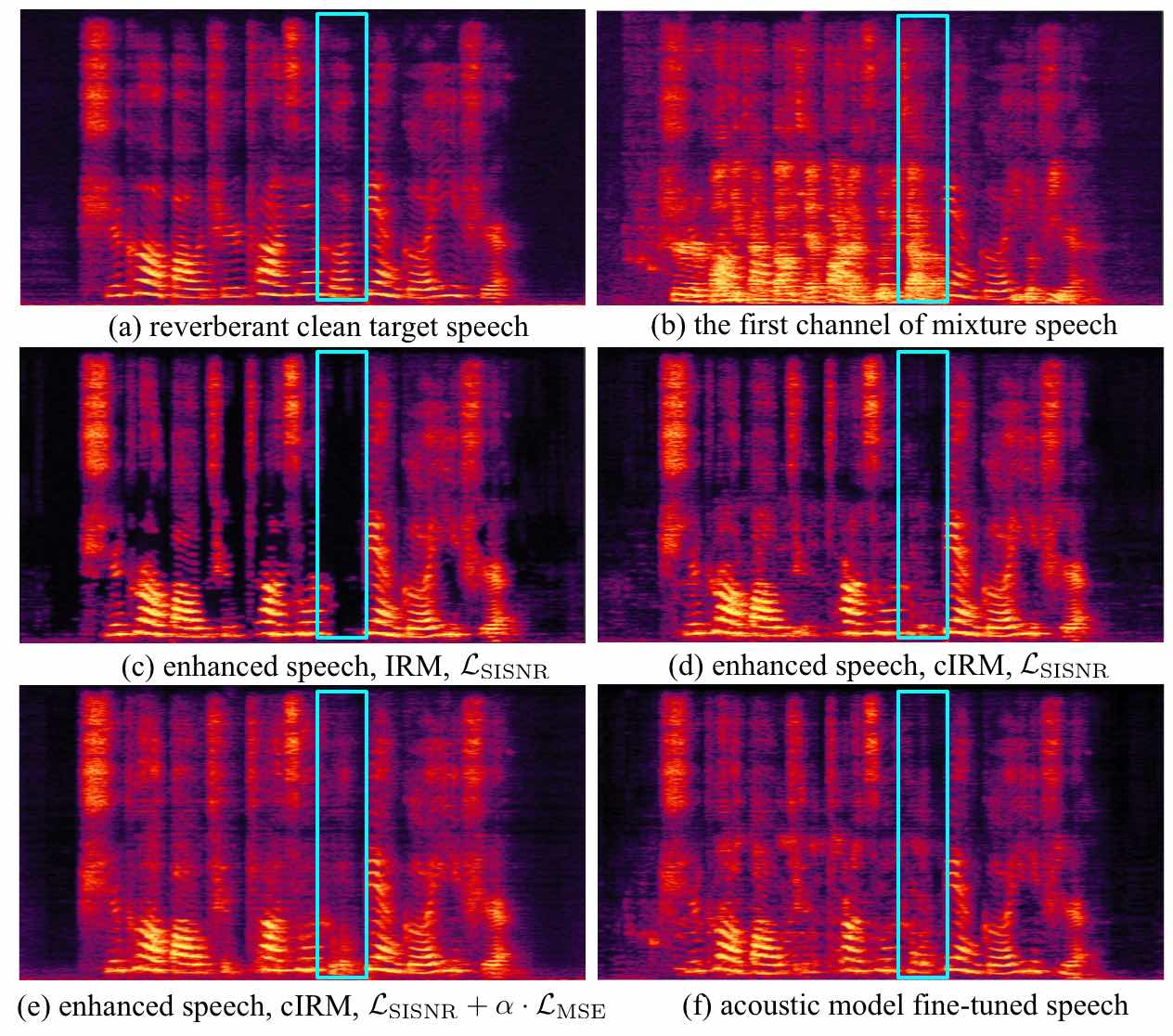}
        \caption{Reverberant, mixture and enhanced spectrograms.}
        \label{fig:spectrogram}    
\end{figure}

Although multi-channel target speech enhancement has been proved effective in terms of PESQ and SDR \cite{bahmaninezhad2019comprehensive,8461639}, directly passing the enhanced signal to ASR systems does not achieve expected improvements in recognition accuracy.
Figure~\ref{fig:spectrogram}~(a) and (b) display spectrograms of a reverberant target speech and the overlapped speech, respectively. For overlapped speech separation, the speech distortion problem is mainly caused by an enhancement algorithm which performs too aggressively, especially when the interfering speakers are stronger than the target speaker. Figure~\ref{fig:spectrogram}~(c) presents the output spectrogram estimated by the IRM-based multi-channel model described in Section \ref{sec:modules} using SI-SNR loss. Due to its destructive interference suppression, lots of holes appear in the enhanced spectrogram when compared with reverberant target speech in (a). The situation is particularly worse in the blue box where an interfering speaker dominates in time-frequency bins and most of the spectrogram contents are removed in the output. The harm of processing artifacts introduced during target speech enhancement may outweigh the benefit brought by interference suppression. We next investigate three types of methods to reduce speech distortions.

\subsection{cIRM-Based Waveform Reconstruction}
The range of $\hat{M}_{\text{IRM}}$ in Section \ref{sec:modules} is $[0,+\infty)$ with ReLu activation, causing the model outputs easily trapped around 0 when the energy of the target speech is lower than the interfering speech, and thus introducing holes in the enhanced spectrogram $\hat{M}_{\text{IRM}}\otimes Y_{1}$. The problem becomes more severe in an sigmoid mask. One way to deal with this drawback is to predict linear cIRM of unbounded range instead:
\vspace{-0.2cm}
\begin{equation}
\hat{M}_{\text{cIRM}}=\hat{M_r}+i\hat{M_i}
\vspace{-0.2cm}
\end{equation}
where $\hat{M_r}, \hat{M_i}\in\left(-\infty,+\infty\right)$. Figure~\ref{fig:spectrogram}~(d) illustrates the enhancement result estimated by a cIRM-based multi-channel model subject to a SISNR loss. Those spectrogram holes apparently disappear when compared with the IRM-based enhanced spectrogram. And most of the enhanced spectrogram at low and intermediate frequencies are restored in the blue box. Studies in \cite{williamson2016complex,williamson2015complex} conclude that the gain of using cIRM is that both the magnitude and phase spectra are jointly estimated in the complex domain. And we observe that the infinite range of cIRM leads to eliminate speech distortions in enhancement. 

\subsection{Objective on Fbank Domain}
Another way to obtain a distortionless spectrogram is to train the enhancement model in a multitask manner with an extra training loss in fbank domain as:
\vspace{-0.2cm}
\begin{equation}
\mathcal{L}=\mathcal{L}_{\text{SISNR}}+\alpha\cdot\mathcal{L}_{\text{MSE}}\{\text{LFB}(\hat{\mathbf{s}})-\text{LFB}(\mathbf{s})\}
\vspace{-0.1cm}
\end{equation}
where LFB($\cdot$) operation extracts log filterbank (LFB) features of estimated waveform $\hat{\mathbf{s}}$ and reverberant target speech $\mathbf{s}$, respectively. $\alpha$ is the weight applied to the loss on LFB domain. By incorporating the loss on LFB domain, the enhancement model tends to predict the target speech whose LFB feature better fits the acoustic model. The blue box in Figure \ref{fig:spectrogram}~(e) shows that high-frequency contents of the enhanced spectrogram are restored. Due to LFB's low resolution in frequency, the overall enhancement is smooth and the harmonics in the blue box turn to be blurred.

\subsection{Acoustic Model Driven Enhancement}
An integrated end-to-end paradigm by jointly modeling the front-end enhancement and back-end acoustic model \cite{wu2017end,bahdanau2016end} is a desirable solution for eliminating the impact of distortion on recognition, since the target speech enhancement front-end is directly optimized towards improved the speech recognition accuracy. A hybrid deep learning framework is adopted to perform the joint training for multi-channel overlapped speech recognition. We directly stack the LFB extraction layer of a convolutional, long short-term memory and fully connected deep neural network (CLDNN) acoustic model on top of the enhancement network's decoder layer. The connectionist temporal classification (CTC) object function used to train the acoustic model is utilized to fine-tune the weights of enhancement and recognition models. The blue box of Figure~\ref{fig:spectrogram}~(f) highlights that acoustic model fine-tuned enhanced speech retains spectral details and observable harmonics, which is noted to be a closer match to the original target speech spectrogram. The gain of joint model is simply summarized from a joint optimization of the speech enhancement and recognition networks in \cite{wu2017end,bahdanau2016end}. Our detailed observation shows that the acoustic model driven enhancement creates less distortion in the output spectrogram, beneficial for recognition. The effectiveness of the above distortionless methods on recognition is demonstrated in Section \ref{sec:exp}.


\section{Experiments}
\label{sec:exp}
\vspace{-0.2cm}
\subsection{Dataset}
\label{sec:dataset}
We simulated a multi-channel reverberant version of two-speaker mixture data set by AISHELL-1 corpus, which is a public data set for Mandarin speech recognition \cite{bu2017aishell}. A 6-element uniform circular array is used as the signal receiver, the radius of which is 0.035~m. The target speaker is mixed with an interfering speaker randomly at signal to interference ratio (SIR) -6, 0 or 6~dB. The classic image method \cite{ISM} is used to add multi-channel room impulse response (RIR) to each source in the mixture and reverberation time (RT60) ranges from 0.05 to 0.5~s. The room configuration (length-width-height) is randomly sampled from 3-3-2.5~m to 8-10-6~m. The microphone array and speakers are at least 0.3~m away from the wall. The distance between microphone array and speakers ranges from 1~m to 5~m. The speaker's direction-of-arrival ranges from 0 to \ang{360}, so that our data set contains samples with the angle difference of two speakers ranging from 0 to \ang{180}. Moreover, the train, validation and test sets consist of 340, 40 and 20 speakers, respectively. The speakers in the three sets are not overlapped, which means approaches are evaluated under speaker-independent scenario. All data is sampled at 16~kHz.

\subsection{System setup}

\subsubsection{Multi-channel target speech enhancement model}
\vspace{-0.2cm}
For the encoder and decoder settings, the kernel size and stride are 512 and 256 samples, respectively. The kernel weights are set according to STFT/iSTFT operation. 257-dimensional LPS feature is extracted based on the output of STFT kernel from the first channel mixture. 6 IPDs are extracted between microphone pairs (1, 4), (2, 5), (3, 6), (1, 2), (3, 4) and (5, 6). Note that, to eliminate the impact of direction of arrival estimation error on our findings, the target speaker's direction is assumed to be known for computing AF.

\subsubsection{CLDNN acoustic model}
\vspace{-0.2cm}
A linear connection layer with 257-dimensional input and 40-dimensional output is used to extract LFB feature from single-channel waveforms with 25-ms window length and 10-ms hop size. The CLDNN model starts with two convolutional layers and then four LSTM layers, each with 512 hidden units, and then two full-connection linear layers plus a softmax layer. We use context-independent phonemes as the modeling units, which form 218 classes in our Chinese ASR system. A tri-gram language model (LM) estimated on AISHELL-1 text is used.

\subsection{Distortionless methods for recognition}

\subsubsection{IRM-Based Waveform Reconstruction (``base'')}
\label{sec:base}
\vspace{-0.2cm}
First, to obtain the best recognition result on test data enhanced by the IRM-based network, we train the acoustic models using different training data and evaluate them on 4 test sets in Table \ref{table:separated}. ``cln.'', ``rev.'', ``mix.'' denote dry clean signal of target speaker, reverberant signal of target speaker and first channel of input signal, respectively. The IRM-based enhancement network using SISNR loss, denoted as ``base'', infers a single-channel output signal ``base-enh'' based on 6-channel overlapped noisy speech. A competitive character error rate (CER) of 11.97\% is attained on clean set ``cln.'' by acoustic model A1 trained on clean set only. Initialized with A1, several multi-condition acoustic models are investigated. Based on all multi-condition training sets,  acoustic model A4 achieves the best performance with a CER of 29.15\%.

\begin{table}[htbp]
\scriptsize
\vspace{-0.1cm}
\caption{CER of clean and multi-condition acoustic models}
\vspace{-0.3cm}
\label{table:separated}
\setlength{\tabcolsep}{0.8mm}
\centering
\begin{tabular}{c|c|cccc|cccc}\toprule
\multicolumn{1}{c|}{\multirow{2}{*}{\textbf{AM}}} & \multicolumn{1}{c|}{\multirow{2}{*}{\textbf{initialized}}} &  \multicolumn{4}{c|}{\textbf{training data}} & \multicolumn{4}{c}{\textbf{test data}}  \\ \cline{3-10}
\multicolumn{1}{c|}{} &\multicolumn{1}{c|}{} & \multicolumn{1}{c}{cln.}  & \multicolumn{1}{c}{rev.} & \multicolumn{1}{c}{mix.}  & \multicolumn{1}{c|}{base-enh}  & \multicolumn{1}{c}{cln.}  & \multicolumn{1}{c}{rev.} & \multicolumn{1}{c}{mix.}  & \multicolumn{1}{c}{base-enh} \\ \hline
 A1 & $\times$ & $\checkmark$ & $\times$  & $\times$ & $\times$& 11.97 & 17.70 & 88.70 &38.66 \\
 A2 & A1 & $\checkmark$ & $\checkmark$ & $\checkmark$  & $\times$&13.25 & 16.16 & 76.20 &37.40 \\
 A3 & A1 & $\times$  & $\times$ & $\times$ &$\checkmark$ & 16.76  & 19.75 & 91.68 & 30.69 \\
 A4 & A1 &$\checkmark$ & $\checkmark$ & $\checkmark$  &$\checkmark$ & 13.76  & 16.21& 83.80 & \textbf{29.15} \\
\bottomrule
\end{tabular}
\end{table}

\begin{table*}[t]
\scriptsize
\caption{CER and PESQ of distortionless methods on different SIRs and angle differences}
\vspace{-0.3cm}
\label{table:simu}
\setlength{\tabcolsep}{2.95mm}
\centering
\begin{tabular}{c|c|c|ccc|cccc|c|c}\toprule
\multicolumn{1}{c|}{\multirow{2}{*}{\textbf{system}}} & \multicolumn{1}{c|}{\multirow{2}{*}{\textbf{mask}}} & \multicolumn{1}{c|}{\multirow{2}{*}{\textbf{loss function}}}  & \multicolumn{3}{c|}{\textbf{SIR}} & \multicolumn{4}{c|}{\textbf{angle difference}}& \multicolumn{1}{c|}{\multirow{2}{*}{\textbf{Avg.}}}  & \multicolumn{1}{c}{\multirow{2}{*}{\textbf{PESQ}}}\\ \cline{4-10} 
\multicolumn{1}{c|}{}  & \multicolumn{1}{c|}{}& \multicolumn{1}{c|}{} & \multicolumn{1}{c}{-6~dB} & \multicolumn{1}{c}{0~dB} & \multicolumn{1}{c|}{6~dB}  & \multicolumn{1}{c}{0$^{\circ}$-15$^{\circ}$} & \multicolumn{1}{c}{15$^{\circ}$-45$^{\circ}$} & \multicolumn{1}{c}{45$^{\circ}$-90$^{\circ}$} & \multicolumn{1}{c|}{90$^{\circ}$-180$^{\circ}$} & \multicolumn{1}{c|}{} & \multicolumn{1}{c}{} \\ \hline
base & IRM & SISNR &  36.92 & 27.88 &22.74 & 38.15 & 28.81 & 27.29 & 26.19 &29.15 & 2.72 \\
sept-1 & cIRM& SISNR &  33.92 & 25.55 & 20.46 & 35.93 & 25.91 & 24.92 & 23.70 &26.62 & 2.86 \\
sept-2 & cIRM & SISNR+MSE(LFB) &28.60 & 22.20 & 18.58 & 31.88 & 22.46 & 21.20 & 20.61 & 23.31 & 3.25 \\
joint & cIRM& CTC &27.49 & 21.03 & 17.25 & 31.52 & 21.17 & 19.84 & 19.17& 21.90 & 2.88 \\
\bottomrule                    
\end{tabular}
\end{table*}

\begin{table*}[t]
\caption{CER of distortionless methods on real recorded RIRs using a mismatched DFSMN-based ASR system}
\scriptsize
\vspace{-0.3cm}
\label{table:real}
\setlength{\tabcolsep}{5.57mm}
\centering
\begin{tabular}{c|c|c|ccc|cc|c}\toprule
\multicolumn{1}{c|}{\multirow{2}{*}{\textbf{system}}} & \multicolumn{1}{c|}{\multirow{2}{*}{\textbf{mask}}} & \multicolumn{1}{c|}{\multirow{2}{*}{\textbf{loss function}}} & \multicolumn{3}{c|}{\textbf{SIR}} & \multicolumn{2}{c|}{\textbf{angle difference}}& \multicolumn{1}{c}{\multirow{2}{*}{\textbf{Avg.}}}\\ \cline{4-8} 
\multicolumn{1}{c|}{}  & \multicolumn{1}{c|}{} & \multicolumn{1}{c|}{} & \multicolumn{1}{c}{-6~dB} & \multicolumn{1}{c}{0~dB} & \multicolumn{1}{c|}{6~dB} & \multicolumn{1}{c}{90$^{\circ}$} & \multicolumn{1}{c|}{180$^{\circ}$}  \\ \hline
cln. & NA   & NA & 1.13 & 1.07 & 1.10  & 1.04& 1.19 & 1.10\\
rev. & NA &  NA & 1.43 & 1.35 & 1.41 &  1.26& 1.60 & 1.40 \\
mix. &  NA &  NA & 96.76 & 75.37& 42.81  & 69.28& 75.16 & 71.64 \\ \cline{1-9} 
base & IRM & SISNR & 21.25& 9.14 & 5.32  & 11.37& 12.57 & 11.87  \\
sept-1 & cIRM& SISNR & 19.11 & 8.12 & 4.80  & 10.35& 11.04 & 10.65 \\
sept-2 & cIRM & SISNR+MSE(LFB) & 17.31 & 7.59 & 4.23  & 9.28& 10.20 & 9.68  \\
joint & cIRM & CTC & 14.08 & 6.20 & 3.63  & 7.86& 8.04 & 7.95 \\
\bottomrule               
\end{tabular}
\end{table*}


\vspace{-0.3cm}
\subsubsection{cIRM-Based Waveform Reconstruction (``sept-1'')}
\vspace{-0.15cm}
We next provide the results of a cIRM-based multi-channel target speech enhancement network subject to a SISNR constraint labeled as ``sept-1'' in Table~\ref{table:simu}. Same as the optimal training strategy in Section~\ref{sec:base}, the acoustic model is trained on ``cln.'', ``rev.'',  ``mix.'' and enhanced data by ``sept-1''. We attain a lower CER of 26.62\% on test data enhanced by ``sept-1''. Moreover, ``sept-1'' consistently outperforms ``base'' in all tested SIRs and angle differences, illustrating the effectiveness of using distortionless cIRM-based reconstructed waveforms for recognition. Besides, cIRM-based enhancement also achieves better speech quality with a PESQ value of 2.86, compared to 2.72 in the IRM-based method. 
 
\subsubsection{Objective on Fbank Domain (``sept-2'')}
\vspace{-0.15cm}
``sept-2'' is a multi-channel target speech enhancement model estimating cIRM under SISNR and mean squared error of LFB constraints. $\alpha=1$ achieves the best CER score in our experiments. On test data enhanced by ``sept-2'', with adding recognition feature constraint, the acoustic model trained on all multi-condition data including  ``cln.'', ``rev.'',  ``mix.'' and ``sept-2'' enhanced data further boosts CER to 23.31\% from 26.62\% in ``sept-1''. Moreover, comparing to ``sept-1'', ``sept-2'' achieves better PESQ value of 3.25.

\subsubsection{Acoustic Model Driven Enhancement (``joint'')}
\vspace{-0.15cm}
Finally, we compare the joint model with the above separately trained systems. ``joint'' is initialized with the well-trained front-end enhancement model ``sept-1'' and back-end acoustic model trained on all multi-condition data including ``cln.'', ``rev.'',  ``mix.'' and ``sept-1'' enhanced speech. A lower CER of 21.90\% is achieved. If we force back-end acoustic model in ``joint'' frozen while front-end enhancement model learnable during joint training,  a worse CER of 22.27\% is obtained, demonstrating the superiority of an end-to-end joint model with trainable enhancement and acoustic models. As shown in Table~\ref{table:simu}, ``joint'' illustrates stable performances and consistently outperforms all separately trained systems in all SIRs and angle difference categories. Specifically, a significant CER decrement is achieved from 29.15\% in ``base'' to 21.90\% using ``joint'', showing a relative improvement of about 25\%. It should be noted that pretraining ``joint'' with a better enhancement model ``sept-2'' or training ``joint'' in a multitask manner as SISNR+CTC can further boost the recognition performance in our supplementary experiments. With matched acoustic modeling, we can see that acoustic model fine-tuned speech achieves the highest recognition accuracy when compared with other two kinds of distortionless waveforms. Although ``joint'' achieves a much lower PESQ of 2.88 relative to 3.25 in ``sept-2'', it boosts CER to 21.90\% from 23.31\% in ``sept-2''. This is consistent with our analysis that a large gain in speech quality measured by objective metrics, does not necessarily lead to improved speech recognition performance due to speech distortion in the enhancement stage. 


\vspace{-0.2cm}
\subsection{Evaluation on real situations}
\vspace{-0.2cm}
It is important to evaluate the distortionless methods in real-world conditions. Considering there is no public overlapped real data in a circular uniform array and collecting mixed speech is very time-consuming, we choose to record real RIRs in 6 realistic rooms. The 6-channel RIRs were measured with various loudspeaker and microphone distances of 0.5, 1,~2,~3~and~5~meters and azimuth angles of 0$^{\circ}$, 90$^{\circ}$, 180$^{\circ}$ and 270$^{\circ}$, so that our real test set contains samples with angle differences of 90$^{\circ}$ and 180$^{\circ}$. Note that angle difference equal to 0$^{\circ}$ can not be handled by direction-aware enhancement algorithms, since the target and interfering speakers are in the same direction. The mixed test data is generated in the same manner as in Section~\ref{sec:dataset}. More importantly, if we conclude that the gain of using cIRM, recognition feature constraint and acoustic model fine-tuned enhancement comes from distortionless waveforms, it is necessary to prove the enhanced distortionless speech consistently helps recognition on a mismatched ASR system that unseens the datasets used to train enhancement and acoustic models. We therefore directly evaluate the real recorded speech, enhanced by ``base'', ``sept-1'', ``sept-2'' and acoustic model fine-tuned enhancement systems without retraining, on a well-trained deep feed-forward sequential memory network (DFSMN)-based ASR system \cite{you2019dfsmn} in Table \ref{table:real}. The standalone ASR system is trained on a 10K-hour mixed Mandarin dataset in application domains and we refer the readers to \cite{you2019dfsmn} for more details. Experimental results show that CERs of 1.10\% and 1.40\% are obtained on clean and reverberant speech, respectively, demonstrating the excellent performance of the DFSMN-based ASR system. Relative to ``base'', all separately and jointly trained systems perform better and the acoustic model fine-tuned enhancement still attains the best CER in all tested SIRs and angle differences.


\vspace{-0.3cm}
\section{Conclusions}\label{con}
We assess three main approaches towards distortionless waveforms for overlapped speech recognition in this paper. We show that all the methods are effective to reduce speech distortions and can improve recognition. And acoustic model fine-tuned enhancement outperforms all separately trained systems for simulated test data with joint-trained acoustic model, or real test data with well-trained standalone acoustic model.


\bibliographystyle{IEEEbib}
\bibliography{main}
\end{document}